\documentclass[12pt]{article}
\usepackage{graphicx}

\def\Re{{\cal R \mskip-4mu \lower.1ex \hbox{\it e}\,}}
\def\Im{{\cal I \mskip-5mu \lower.1ex \hbox{\it m}\,}}
\def\ie{{\it i.e.}}

\def\etal{{\it et al.}}

\def\sub#1{_{\lower.25ex\hbox{$\scriptstyle#1$}}}
\def\tev{\,{\ifmmode\mathrm {TeV}\else TeV\fi}}
\def\gev{\,{\ifmmode\mathrm {GeV}\else GeV\fi}}
\def\mev{\,{\ifmmode\mathrm {MeV}\else MeV\fi}}
\def\mpl{\ifmmode M_{*}\else $M_{*}$\fi}
\def\to{\rightarrow}

\def\subw{_{\rm w}}
\def\mh{\ifmmode m\sbl H \else $m\sbl H$\fi}
\def\mch{\ifmmode m_{H^\pm} \else $m_{H^\pm}$\fi}
\def\mt{\ifmmode m_t\else $m_t$\fi}
\def\mc{\ifmmode m_c\else $m_c$\fi}
\def\mz{\ifmmode M_Z\else $M_Z$\fi}
\def\mw{\ifmmode M_W\else $M_W$\fi}
\def\mws{\ifmmode M_W^2 \else $M_W^2$\fi}
\def\mhs{\ifmmode m_H^2 \else $m_H^2$\fi}   
\def\mzs{\ifmmode M_Z^2 \else $M_Z^2$\fi}
\def\mts{\ifmmode m_t^2 \else $m_t^2$\fi}
\def\mcs{\ifmmode m_c^2 \else $m_c^2$\fi}
\def\mchs{\ifmmode m_{H^\pm}^2 \else $m_{H^\pm}^2$\fi}
\def\ztwo{\ifmmode Z_2\else $Z_2$\fi}
\def\zone{\ifmmode Z_1\else $Z_1$\fi}
\def\mtwo{\ifmmode M_2\else $M_2$\fi}
\def\mone{\ifmmode M_1\else $M_1$\fi}
\def\tb{\ifmmode \tan\beta \else $\tan\beta$\fi}
\def\xw{\ifmmode x\subw\else $x\subw$\fi}
\def\ch{\ifmmode H^\pm \else $H^\pm$\fi}
\def\lum{\ifmmode {\cal L}\else ${\cal L}$\fi}
\def\inpb{\,{\ifmmode {\mathrm {pb}}^{-1}\else ${\mathrm {pb}}^{-1}$\fi}}
\def\infb{\,{\ifmmode {\mathrm {fb}}^{-1}\else ${\mathrm {fb}}^{-1}$\fi}}
\def\epem{\ifmmode e^+e^-\else $e^+e^-$\fi}
\def\ppb{\ifmmode \bar pp\else $\bar pp$\fi}
\def\bsg{\ifmmode B\to X_s\gamma\else $B\to X_s\gamma$\fi}
\def\bsll{\ifmmode B\to X_s\ell^+\ell^-\else $B\to X_s\ell^+\ell^-$\fi}
\def\bstt{\ifmmode B\to X_s\tau^+\tau^-\else $B\to X_s\tau^+\tau^-$\fi}
\def\lamt{\ifmmode \tilde\lambda\else $\tilde\lambda$\fi}
\def\shat{\ifmmode \hat s\else $\hat s$\fi}
\def\that{\ifmmode \hat t\else $\hat t$\fi}
\def\uhat{\ifmmode \hat u\else $\hat u$\fi}

\newskip\zatskip \zatskip=0pt plus0pt minus0pt
\def\matth{\mathsurround=0pt}
\def\lsim{\mathrel{\mathpalette\atversim<}}

\def\atversim#1#2{\lower0.7ex\vbox{\baselineskip\zatskip\lineskip\zatskip
  \lineskiplimit 0pt\ialign{$\matth#1\hfil##\hfil$\crcr#2\crcr\sim\crcr}}}

\renewcommand{\thefootnote}{\fnsymbol{footnote}}

\begin{document} \begin{titlepage} 
\rightline{\vbox{\halign{&#\hfil\cr
&SLAC-PUB-9127\cr
&January 2002\cr}}}
\begin{center}

{\Large\bf Black Hole Production at the LHC: Effects of Voloshin Suppression}
\footnote{Work supported by the Department of 
Energy, Contract DE-AC03-76SF00515}
\medskip

\normalsize 
{\bf \large Thomas G. Rizzo}
\vskip .3cm
Stanford Linear Accelerator Center \\
Stanford University \\
Stanford CA 94309, USA\\
\vskip .2cm

\end{center}

\begin{abstract} 
We examine the rates for the production of black holes(BH) at the LHC 
in light of the exponential suppression of the geometric cross section estimate 
recently proposed by Voloshin. We show that the resulting production rates 
will still be quite large over a reasonably wide range of model parameters. 
While BH production may no longer be the dominant collider process, its unique 
signature will ensure observability over more conventional backgrounds. 
\end{abstract} 




\renewcommand{\thefootnote}{\arabic{footnote}} \end{titlepage}


Theories with extra dimensions and a low effective Planck scale($\mpl$) offer 
the exciting possibility that the production rate of black holes(BH) somewhat 
more massive than $\mpl$ can be quite large at future colliders. For example, 
cross sections of order 100 pb at the LHC{\cite {gids}}, and even 
larger ones at the VLHC, have been advertised in the analyses presented by 
Giddings and Thomas(GT) and by Dimopoulos and Landsberg(DL). Although 
in practice 
the actual production cross section critically depends on the BH mass, 
the exact value of $\mpl$ and the number of extra dimensions, following the 
analysis of the authors in Ref.{\cite {gids}}, one 
finds very large rates over almost all of the interesting parameter space. 
These earlier analyses and discussions of the production of BH at colliders 
have been elaborated upon by several groups of authors{\cite {sgid}} and the 
production of BH by cosmic rays has also been considered{\cite {cosmic}}. The 
most important question to address is whether or not the BH cross sections are 
actually this large or, at the very least, large enough to lead to visible 
rates at future colliders. 

The basic idea behind the original collider BH papers is as follows: we 
consider the collision of two high energy Standard Model(SM) partons which are 
confined to a 3-brane, as they are in 
both the models of Arkani-Hamed, Dimopoulos and Dvali(ADD){\cite {add}} and 
Randall and Sundrum(RS){\cite {rs}}. 
In addition, we imagine that gravity is free to propagate in 
$\delta$ extra dimensions with the $4+\delta$ dimensional Planck scale 
assumed to be $\mpl \sim 1$ TeV. The curvature of the space is assumed to be 
small compared to the energy scales involved in the collision process 
so that quantum 
gravity effects can be neglected. When these partons have a center of 
mass energy in excess of $\sim \mpl$ and the impact parameter for the 
collision is less than the 
Schwarzschild radius, $R_S$, associated with this center of mass energy, a 
$4+\delta$-dimensional BH is formed with reasonably high efficiency. It is 
expected that a very large fraction of the collision energy, \ie, $\sqrt s$, 
actually goes into the BH formation process so that $M_{BH}\simeq \sqrt s$. 
The subprocess cross section for the production of a non-spinning BH 
is thus essentially geometric for {\it each} pair of initial partons: 
\begin{equation}
\hat \sigma \simeq \epsilon \pi R_S^2\,;
\end{equation}
where $\epsilon$ is an factor that accounts for finite impact parameter and 
angular momentum corrections and which is expected to be $\lsim 1$. (We 
will assume that $\epsilon=1$ in our calculations below.) 
We note that the $4+\delta$-dimensional Schwarzschild radius, $R_S$, scales as
\begin{equation} 
R_S \sim \Big[{M_{BH}\over {\mpl^{~2+\delta}}}\Big]^{1\over {1+\delta}}\,,
\end{equation}
apart from an overall $\delta$- and {\it convention-dependent} numerical 
prefactor. This convention 
dependence has been found to be somewhat confusing for unprepared 
readers of the literature and particularly to experimenters investigating 
the BH production phenomena. The confusion can be traced back to two 
different expressions for the $4+\delta$ dimensional 
Schwarzschild radius used by the original two sets of authors GT and DL. 
Explicitly, the two sets of authors employed distinct relationships 
between their $4+\delta$-dimensional Planck masses, $M_{GT,DL}$, and 
the associated generalized Newton's constant, $G_{4+\delta}$: one finds that 
$M_{DL}^{2+\delta}=G_{4+\delta}^{-1}$, while 
$M_{GT}^{2+\delta}=(2\pi)^\delta /4\pi G_{4+\delta}$. 
Depending on how the input 
parameters are chosen, \ie, which of the above masses is identified with 
$M_*$, this scheme dependent 
prefactor can turn out to be relatively important. In particular, it is found 
to  lead to an apparently very different $\delta$-dependence for 
the BH production cross section in the two cases. In the DL case the 
$\delta$-dependence of the numerical prefactor is rather weak whereas is it 
somewhat stronger in the GT analysis. Naively, for the same input value of 
$M_{BH}$ one finds the ratio of the cross sections obtained by 
the two sets of authors to be 
\begin{equation}
{\hat \sigma_{GT}\over {\hat \sigma_{DL}}}=\Big[{(2\pi)^\delta \over {(4\pi)
}}\Big]^{2\over {1+\delta}}~\Big[{M_{DL}^2\over {M_{GT}^2}}\Big]^{{2+\delta}
\over {1+\delta}}\,,
\end{equation}
which is always greater than unity for $\delta \geq 2$ and grows as $\delta$ 
increases {\it if} one assumes $\mpl=M_{GT}=M_{DL}$ as an input. (The later 
equality is, of course, invalid.) It is 
important to note, however, that when the 
differences in the definitions of the Planck scale are accounted for, both 
cross sections lead to the {\it same numerical results}. This can be 
accomplished by using a common value of $G_{4+\delta}$ as input in both cases. 
Written in terms of $G_{4+\delta}$ the above ratio of cross sections is unity. 
Unfortunately, since $G_{4+\delta}$ is a rather cumbersome object to use as 
an input parameter, in our 
discussion below we will display results for both sets of notation, \ie, 
we will identify $M_*$ as {\it either} $M_{DL}$ or $M_{GT}$.

\begin{figure}[htbp]
\centerline{
\includegraphics[width=9cm,angle=90]{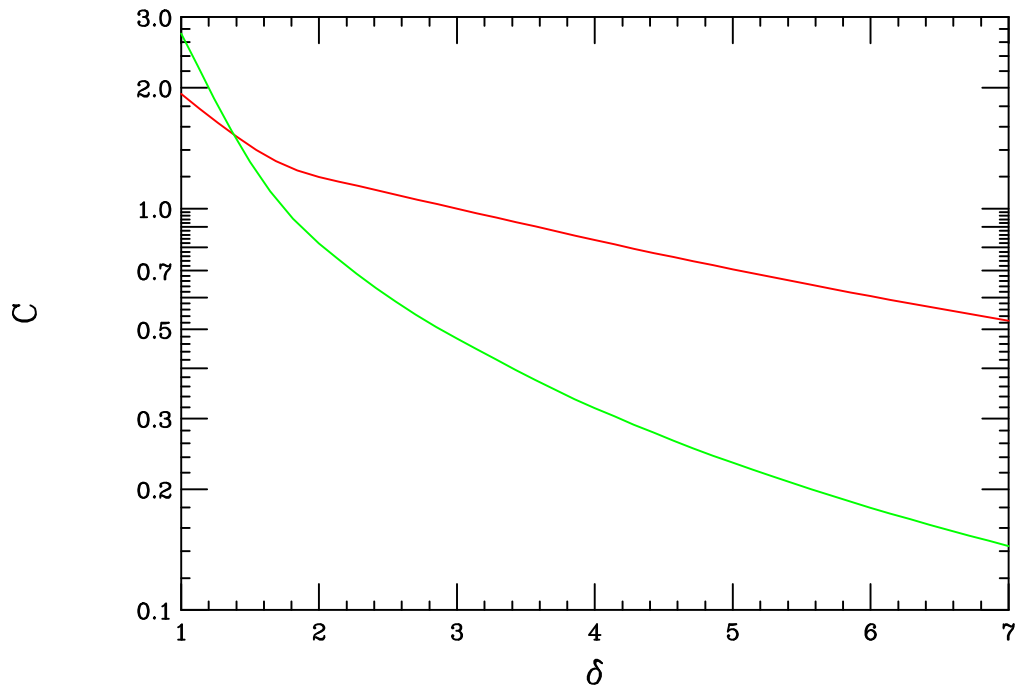}}
\vspace*{0.1cm}
\caption{The parameter $C$ appearing in the exponential suppression factor of 
Voloshin as a function of the number of extra dimensions, 
$\delta$. The upper(lower) curve on the right-hand side of the plot is for 
the GT(DL) scheme.}
\label{cplot}
\end{figure}

The approximate geometric subprocess cross section expression is claimed to 
hold by both GT and DL when the ratio $M_{BH}/\mpl$ is ``large", \ie, 
when the system can be treated semi-classically and quantum gravitational 
effects are small; one may debate just what ``large" really means, but it 
most likely means ``at least a few". Certainly when $M_{BH}/\mpl$ is near 
unity one might expect curvature and stringy effects to become important and 
even the finite 
extent of the incoming partons associated with this stringy-ness would need 
to be considered. Clearly caution must be applied when $M_{BH}\simeq \mpl$ 
in interpreting cross sections evaluated in this parameter space region. 

In order to obtain the actual cross section at 
a collider, one takes the geometric parton-level result, folds in the 
appropriate parton densities, sums over all pairs of possible partons and then 
integrates 
over the relevant kinematic variables. The resulting total cross section for 
BH with masses $\geq M_{BH}^{min}$ is then given by the expression  
\begin{equation}
\sigma=\int^1_{M_{BH}^{min~2}/s} d\tau \int^1_\tau {dx\over {x}} \sum_{ab} 
f_a(x) f_b(\tau /x) ~\hat \sigma(M_{BH})\,, 
\end{equation}
where we have summed over all possible pairs of initial state partons with 
their associated densities $f_i(x)$. These parton densities are evaluated at a 
scale $Q^2=M_{BH}^2$ in the numerical results presented below. 

\begin{figure}[htbp]
\centerline{
\includegraphics[width=9cm,angle=90]{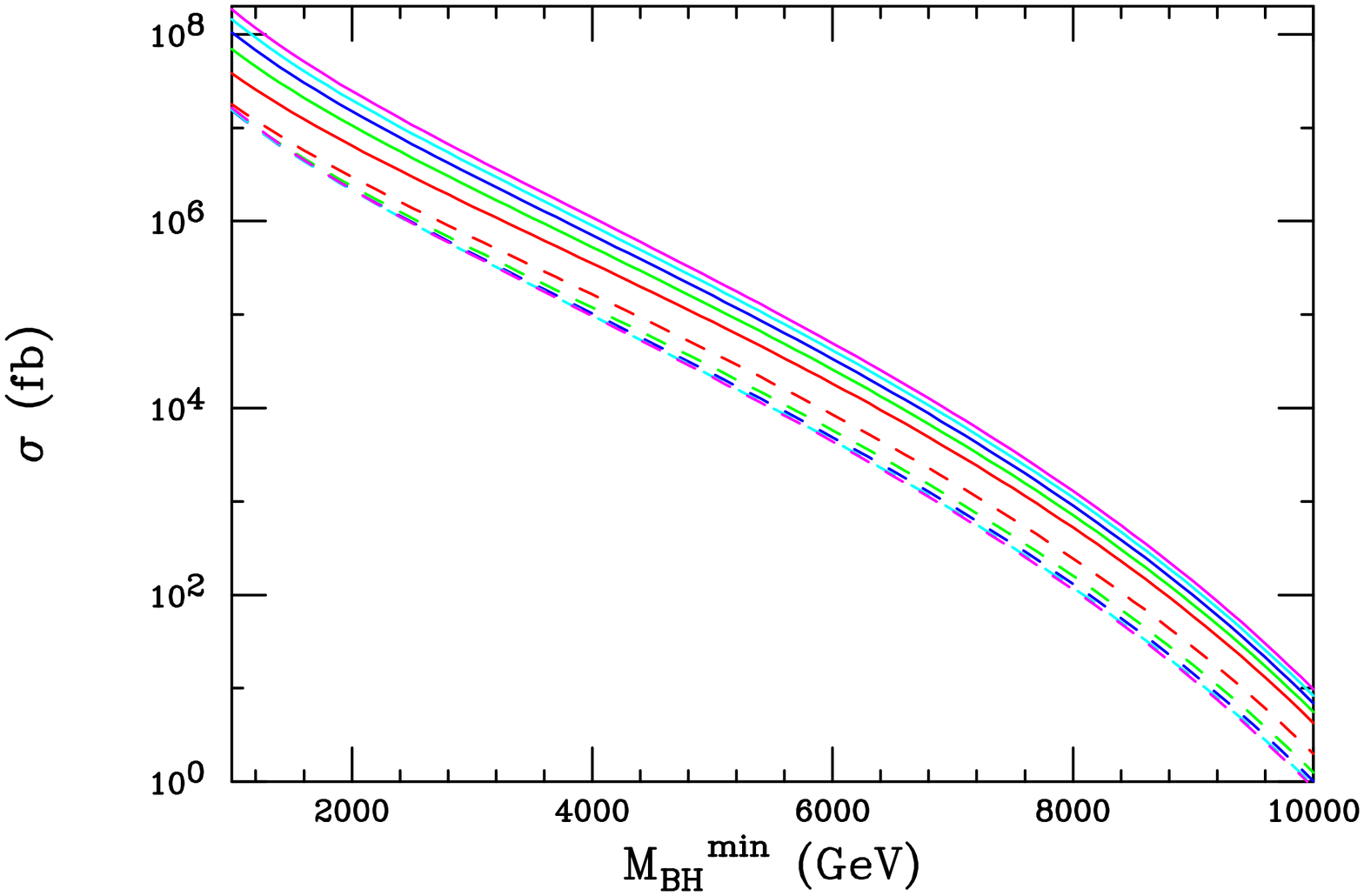}}
\vspace*{5mm}
\centerline{
\includegraphics[width=9cm,angle=90]{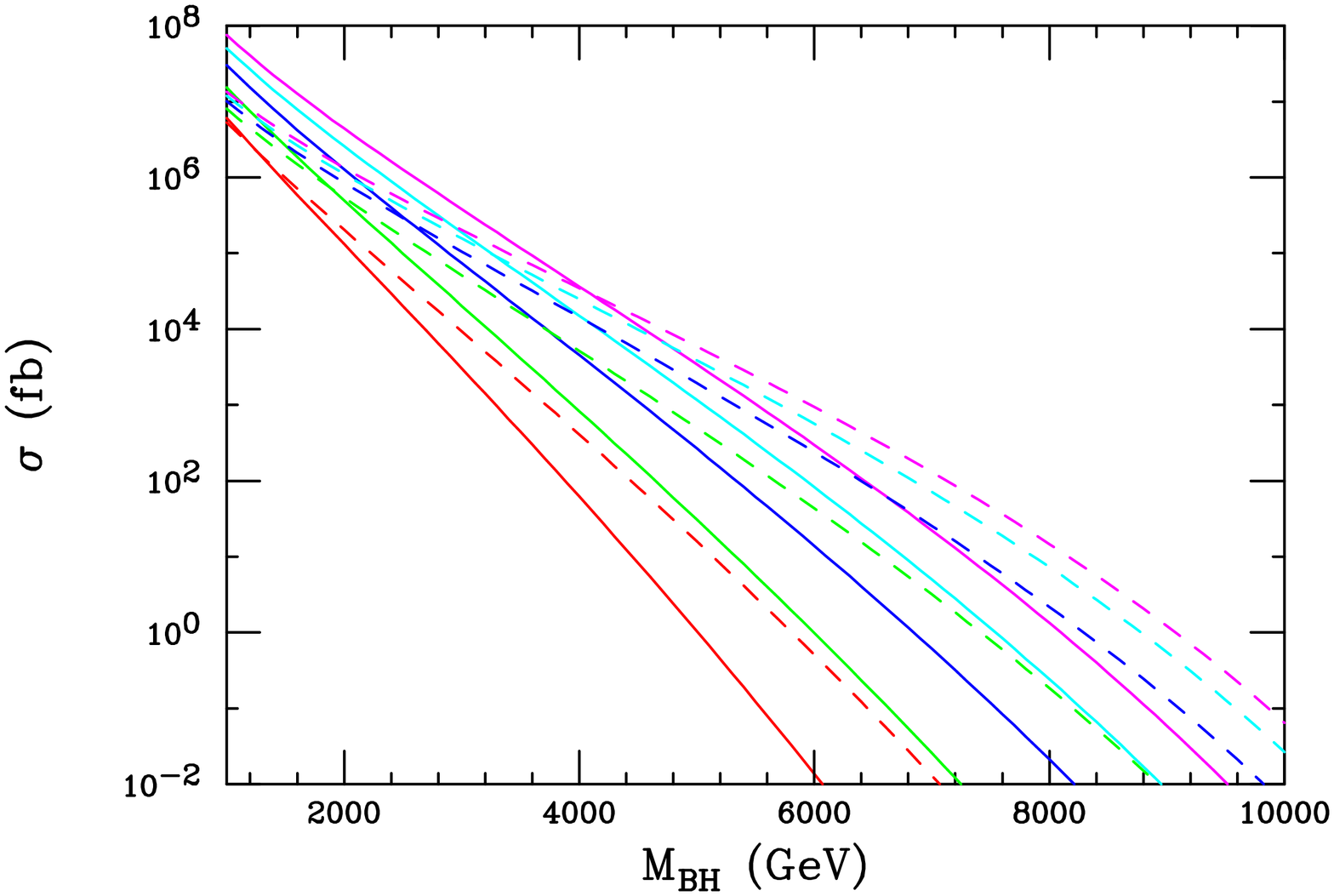}}
\vspace*{0.1cm}
\caption{(Top)Cross section for the production of BH more massive 
than $M_{BH}^{min}$ at the LHC assuming $M_* =1$ TeV  for $\delta=2(3,4,5,6)$ 
extra dimensions with the solid(dashed) curves showing the results 
by Giddings and Thomas(Dimopoulos and Landsberg). In the GT(DL) case an 
increase in the number of extra dimensions leads to an increase(decrease) of 
the cross section. We identify $M_*$ with either $M_{GT}$ or 
$M_{DL}$ for the appropriate set of curves. (Bottom)Same 
as the Top, but now including the effects of the Voloshin damping factor. 
We observe that large cross sections remain possible over a reasonable range of 
the model parameters.  In both cases, DL or GT, the cross section now 
increases with the number of extra dimensions.}
\label{p3-33_snowholes}
\end{figure}

Voloshin has recently{\cite {voloshin}} provided several arguments which 
suggest that an additional exponential suppression 
factor, $S=e^{-I_E}$, must be included which seriously damps the 
pure geometric cross section for this process even in 
the semi-classical case. This implies we should rescale the purely 
geometric result 
as $\hat \sigma \to S\hat \sigma$ in the equations above. Here, $I_E$ is the 
Euclidean Gibbons-Hawking action for the BH{\cite {GH}} generalized to the 
case of $4+\delta$ dimensions. Then $S$ is given explicitly by    
\begin{equation}
S=exp[-4\pi R_S M_{BH}/(1+\delta)(2+\delta)] = exp \Big[-C
\Big({M_{BH}\over {\mpl}}\Big)^{{2+\delta}\over {1+\delta}}\Big]\,, 
\end{equation}
where $C$ is a relatively small, though $\delta$-dependent, constant whose 
detailed form is 
also dependent upon whether the GT or DL schemes are followed. 
Fig.~\ref{cplot} shows 
the value of $C$ as a function of $\delta$ for both schemes. In the DL(GT) 
scheme, for example, when $\delta=3$ one obtains $C=0.474(0.999)$; similarly 
for $\delta=5$ one obtains $C=0.233(0.706)$. Note that in either scheme we 
see that $C$ is rather small and 
decreases substantially as $\delta$ is increased. Again, when the 
differences in notation in the two schemes are accounted for , \ie, using the 
same value of $G_{4+\delta}$, the value of 
the suppression factor is identical. The smallness of $C$ thus seriously 
reduces the numerical impact of the exponential suppression. 
From the $\delta$-dependence of $C$ and $\delta$-dependence of 
the power of the ratio $M_{BH}/M_*$ in the exponential,  
it is clear that the Voloshin suppression will become less effective as the 
number of extra dimensions increases.

While the possibility of the existence of exponential suppression 
remains controversial, and strong arguments have 
been made on either side of the argument, for purposes of this discussion we 
will assume that it is indeed present. (However, we warn the reader 
that the jury is still out on this issue. In either case we anxiously await 
the resolution of this important argument.) If Voloshin's criticisms of the 
geometrical cross section are valid one may certainly worry that the resulting 
exponentially suppressed rates for heavy BH production 
will possibly be too small to be observable at the LHC; as we will see below 
this need not be so. This is partly due to the smallness of $C$ and partly 
due to the expected huge size of the unsuppressed production rates for BH at 
the LHC.
 
Just how do the suppressed and unsuppressed cross sections at the LHC compare? 
As can be seen in Fig.~\ref{p3-33_snowholes} for the case $\mpl=1$ TeV, 
the unsuppressed rates for BH production at 
the LHC are quite large over a wide range of masses and numbers of extra 
dimensions using either set of authors' cross section expressions. (It is 
important to remember that in this 
figure and the others below we appropriately 
identify $\mpl$ as {\it either} $M_{GT}$ or $M_{DL}$ 
depending on which set of predictions are under discussion.) Note that 
the results of Giddings and Thomas always appear larger than those of 
Dimopoulos and Landsberg due to the different definitions used for the Planck 
scale and that these apparent differences between the two sets of 
predictions increases as $\delta$ increases as was discussed above. This 
result is of course just a simple artifact of the known notational difference. 
We also see that Fig.~\ref{p3-33_snowholes} shows the effects of 
the suppression predicted by Voloshin in the two cases.
From these results we make the important observation that for at least 
for some ranges of parameters BH will still be produced at rates that are 
large enough 
to be observable at the LHC {\it even when the Voloshin suppression is active}. 
For example, assuming that $M_{BH}^{min}=5$ TeV, we see that cross sections 
can easily be in the 100-1000 fb range. Although this is not a huge 
cross section the associated rates at the LHC will be quite large given an 
integrated luminosity of order 100 $fb^{-1}$/yr. 
Note that the Voloshin suppression factor apparently 
modifies the two sets of predictions in quite 
different manners due to the two different expressions used for $R_S$. Since 
$(R_S)_{GT}>(R_S)_{DL}$ for all $\delta \geq 2$, assuming the same input 
values for $M_{GT}$ and $M_{DL}$, the GT results are found to appear 
more suppressed than are those of DL. Again, if we take the notational 
differences in the two schemes into account, identical cross sections are 
obtained in both cases.

\begin{figure}[htbp]
\centerline{
\includegraphics[width=9cm,angle=90]{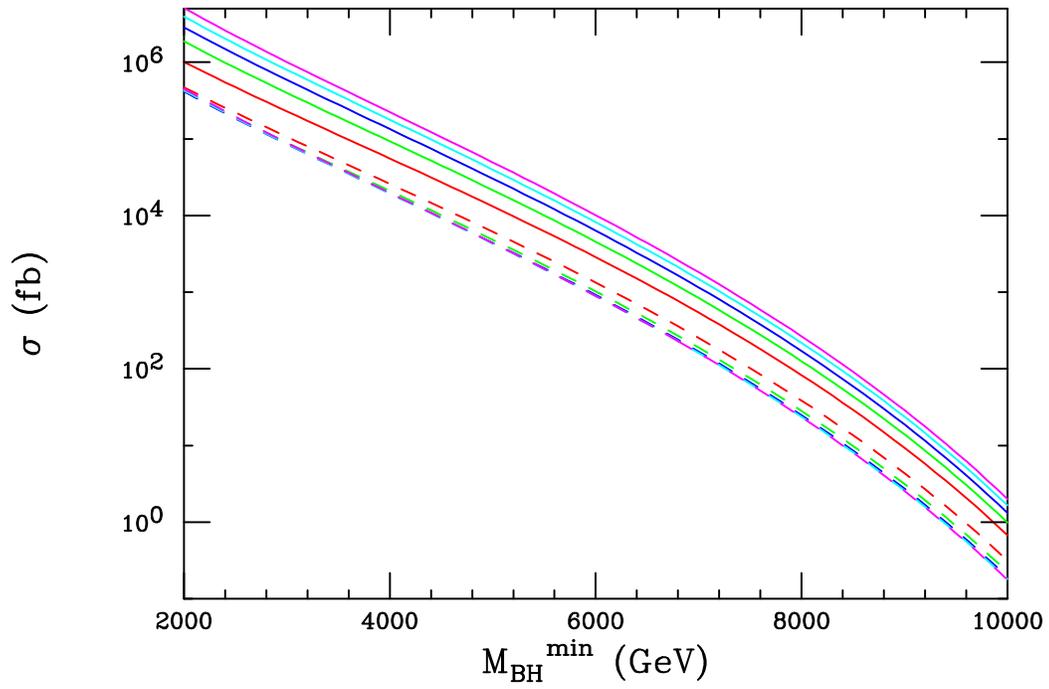}}
\vspace*{15mm}
\centerline{
\includegraphics[width=9cm,angle=90]{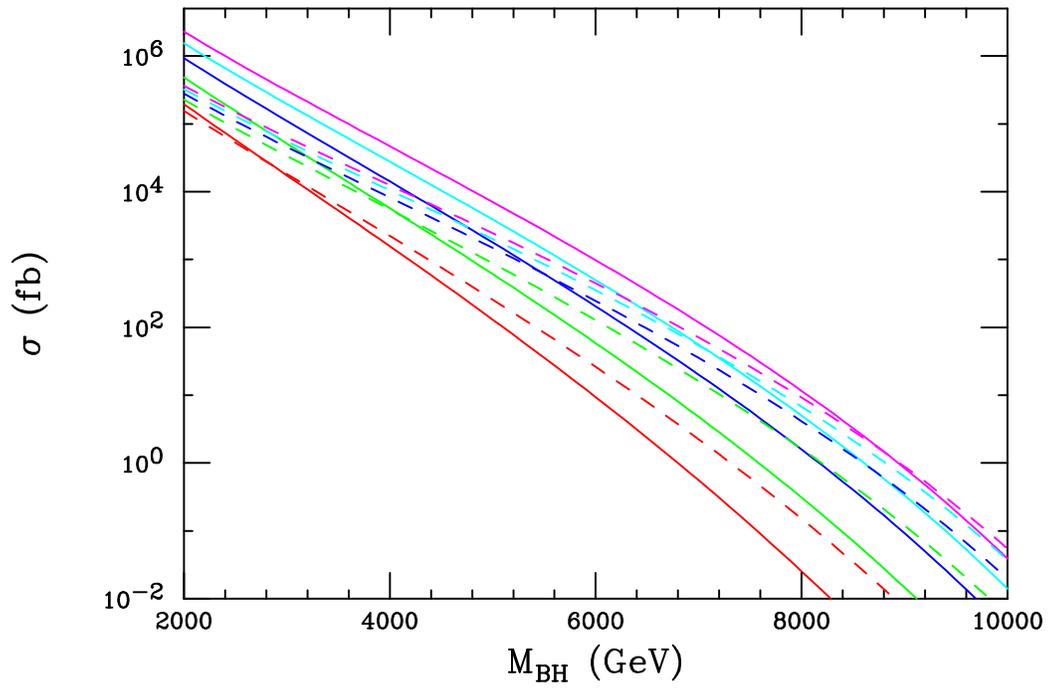}}
\vspace*{0.1cm}
\caption{Same as the last figure but now for a larger value of the 
fundamental Planck scale, $M_* =2$ TeV.}
\label{p3-39_snowholes2}
\end{figure}

\begin{figure}[htbp]
\centerline{
\includegraphics[width=9cm,angle=90]{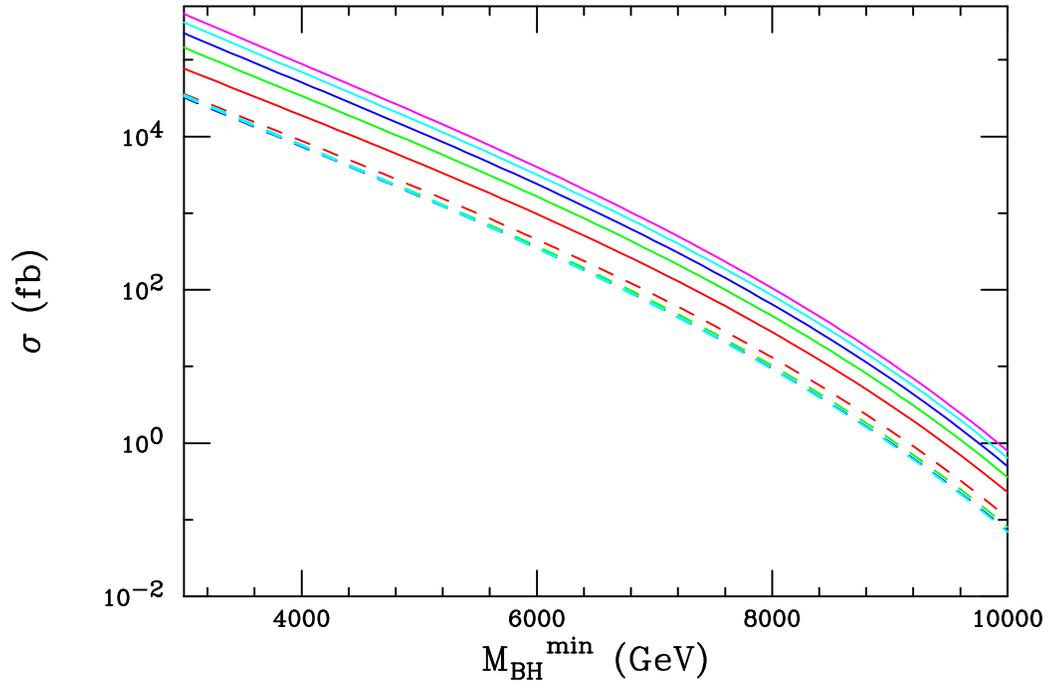}}
\vspace*{15mm}
\centerline{
\includegraphics[width=9cm,angle=90]{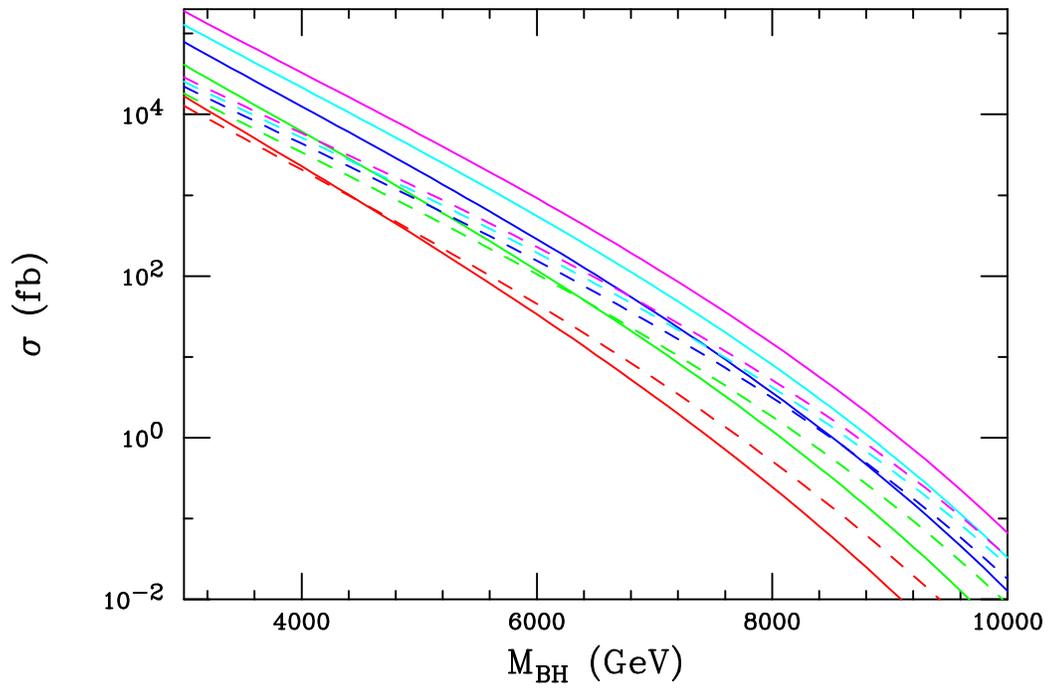}}
\vspace*{0.1cm}
\caption{Same as the last figure but now for $M_* =3$ TeV.}
\label{p3-39_snowholes3}
\end{figure}

What happens as we vary $\mpl$? Figs.~\ref{p3-39_snowholes2} and 
~\ref{p3-39_snowholes3} show the effects of increasing $\mpl$ from 1 TeV to 
2 and 3 TeV, respectively. 
As expected the unsuppressed rates for any fixed value of $M_{BH}$ 
decreases but we also see that the Voloshin suppression becomes less 
effective. This is also to be expected since the ratio $M_{BH}/\mpl$ in the 
exponent of the factor $S$ has been decreased for fixed $M_{BH}$. Again we see 
that for BH in the 5-6 TeV range it is relatively likely that the production 
cross section can quite easily be in excess of 100 fb. 

We remind the reader that once 
produced these BH essentially decay semi-classically, mostly on the brane, 
via Hawking radiation  
into a reasonably large number $\simeq 25$ or more final state partons with 
energies of order 50-100 GeV in a 
highly spherical pattern. Hadrons will be seen to 
dominate over leptons by a factor of 
order 5-10 for such final states. These unusual signatures would not 
be missed at either 
hadron or lepton colliders. (We note that an alternative decay scenario has 
been advocated by Casadio and Harms{\cite {casa}}.) These features are 
sufficiently unique that BH production above conventional backgrounds should 
be observable at the LHC even if the cross sections are substantially smaller 
than the original estimates.

In summary, 
we have examined the production of BH at the LHC assuming that the exponential  
suppression of the geometric cross section predicted by Voloshin is realized.
We have found that even when this suppression is significant the resulting BH 
production 
rates are still quite large for a wide range of model parameters given an 
integrated luminosity of order 100 $fb^{-1}$. If the scale $M_*$ is 
$\sim $ TeV then BH production should provide an exciting signature at the 
LHC and open a door to even more exciting physics.

\noindent{\Large\bf Acknowledgements}

The author would like to thank S. Thomas, G. Landsberg and J. Hewett for 
discussions related to this work.

%
\def\MPL #1 #2 #3 {Mod. Phys. Lett. {\bf#1},\ #2 (#3)}
\def\NPB #1 #2 #3 {Nucl. Phys. {\bf#1},\ #2 (#3)}
\def\PLB #1 #2 #3 {Phys. Lett. {\bf#1},\ #2 (#3)}
\def\PR #1 #2 #3 {Phys. Rep. {\bf#1},\ #2 (#3)}
\def\PRD #1 #2 #3 {Phys. Rev. {\bf#1},\ #2 (#3)}
\def\PRL #1 #2 #3 {Phys. Rev. Lett. {\bf#1},\ #2 (#3)}
\def\RMP #1 #2 #3 {Rev. Mod. Phys. {\bf#1},\ #2 (#3)}
\def\NIM #1 #2 #3 {Nuc. Inst. Meth. {\bf#1},\ #2 (#3)}
\def\ZPC #1 #2 #3 {Z. Phys. {\bf#1},\ #2 (#3)}
\def\EJPC #1 #2 #3 {E. Phys. J. {\bf#1},\ #2 (#3)}
\def\IJMP #1 #2 #3 {Int. J. Mod. Phys. {\bf#1},\ #2 (#3)}
\def\JHEP #1 #2 #3 {J. High En. Phys. {\bf#1},\ #2 (#3)}

\end{document}